\documentstyle{l-aa}

\begin{document}

\thesaurus{02.01.1, 02.04.2, 02.13.2, 02.20.1, 02.23.1, 09.03.2}

\title{Cosmic Ray Momentum Diffusion in Magnetosonic versus
Alfv\'enic Turbulent Fields }
\author{G. Micha{\l}ek \inst{1} , M. Ostrowski\inst{1,2}
and R. Schlickeiser\inst{2} }

\institute{Obserwatorium Astronomiczne, Uniwersytet
 Jagiello\'nski, ul.Orla 171, 30-244 Krak\'ow, Poland
\and
Max-Planck-Institut f\"ur Radioastronomie, Postfach 2024, 53010 Bonn,
Germany}

\offprints{G. Micha\l ek (michalek\@@oa.uj.edu.pl)}

\date{Received ...; accepted .. ; }

\maketitle
\markboth{Cosmic Ray Momentum Diffusion}{in Magnetosonic versus
Alfv\'enic Turbulent Fields}

\begin{abstract}
The acceleration of energetic particle transport in high amplitude magnetosonic
and Alfv\'enic turbulence is considered using the Monte Carlo particle
simulations which involve integration of particle equations of motion.
We derive the momentum diffusion coefficient $D_p$ in the presence of
anisotropic turbulent wave fields in the low-$\beta$ plasma, for
 a flat and a Kolmogorov-type
turbulence spectrum.
 We confirm
the quasilinear result (cf. Schlickeiser \& Miller (1997)) of
enhanced values of $D_p$ due to transit-time damping
resonance interaction in the presence of isotropic fast-mode
waves as compared to the case of slab Alfv\'en waves of the same amplitude. The
relation of $D_p$ to the turbulence amplitude and anisotropy is
investigated.


\end{abstract}

\section{Introduction}
Many astronomical objects (extragalactic radio sources,
supernova remnants, solar flares) emit radiation with  non-thermal spectra.
These  emissions are often connected with the existence of
a hot turbulent magnetized plasma providing conditions for particle
acceleration by MHD turbulence.
 It was first shown by Hall and Sturrock (1967) and by
Kulsrud and Ferrari (1971) that charged particles can be accelerated
by MHD turbulence having wavelengths long compared to the particle gyration
radius. For example, the
importance of a Fermi-like acceleration mechanism
was proved for extragalactic radio sources
(Burn 1975; De Young 1976; Blandford and Rees 1978; Achterberg 1979
and Eilek 1979)  and
for the second phase acceleration in solar flares (Melrose 1974; Ramaty 1979). The interaction between
the waves and particles is determined primarily by resonances
$\omega- k_{\parallel} v_{\parallel} = n \Omega$ between particles
characterized by their parallel velocity $v_{\parallel }$ and gyrofrequency $\Omega $ and
undamped waves characterized by their frequency $\omega $ and wavenumber $k$
\footnote{
The notation is explained in Appendix A}.
At such interactions  energies vary in a  diffusive way, and over a long
timescale a net energy gain results in the process of stochastic
acceleration. For $n \ne 0$, a particle in its parallel rest frame ($v_{\parallel }=0$)
resonates with those waves that it sees at an integral multiple of a gyrofrequency
and for $n=0$ at zero frequency.
Then, a particle feels a net force which is not averaged out by the
phase mixing
(Stix 1962). It is a magnetic analogy of the Landau damping and is
associated with the first-order change in $\delta |\vec{B}|$.
 This type of coupling
 is not observed in the case of Alfv\'en waves but  becomes important
for the magnetosonic waves containing  compressive components of $\delta
\vec{B}$.
The interaction between the particle magnetic momentum  and the parallel
gradient of the magnetic field is called transit-time damping
(cf. discussions by Lee and V\"olk 1975,  Achterberg 1981,
Miller et al. 1996).
Recently, Schlickeiser \& Miller (1997) presented a quasi-linear
derivation of cosmic ray transport coefficients in the presence of MHD
waves, including the isotropic fast-mode turbulence. For the isotropic
Kolmogorov turbulence they demonstrated that the Fokker-Planck
coefficients depend both on the transit-time damping and the gyro-resonance
interactions. For cosmic ray particles with $v \gg V_A$ and a
vanishing turbulence cross helicity the momentum diffusion coefficient
in the fast-mode turbulence is mainly determined by the transit-time
damping contribution, leading to a more effective stochastic acceleration
in comparison to the process in the presence of the pure Alfv\'enic
turbulence.

The  aim of the present paper is to study the momentum diffusion
coefficient $D_p$ in the presence of non-linear ($\delta B \geq B_0$)
magnetosonic and Alfv\'en waves where the quasilinear approximations
do not apply. We also study the influence of the degree of wave propagation
anisotropy and
of the  wave spectral slope on the value of $D_p$.
To consider these problems
the Monte Carlo simulations involving derivations of particles trajectories
in the space filled with finite amplitude fast-mode and Alfv\'en waves are
applied.
Anisotropic wave distributions are modeled by
choosing their wave vectors from finite opening cones directed along
the mean magnetic field $\vec{B}_o$. For our simulations we adopt fast-mode
or Alfv\'en mode turbulence with flat ($q=1$) and
Kolmogorov-type ($q=5/3$) power law
spectrum $\propto k^{-q}$ above the minimum wavenumber $k_{min}$.
We confirm
a substantial increase of $D_p$ for the fast-mode waves in comparison to the
Alfv\'en waves of the same amplitude if the nearly perpendicularly propagating
waves ($\vec{k} \perp
\vec{B}_o$) are included.

\section{Quasi-linear momentum diffusion coefficient}
The quasi-linear theory treats the effect of the weakly perturbed magnetic
field as perturbations of orbits of particles moving in the average
background field. Schlickeiser (1989) has considered
the quasilinear transport
and acceleration parameters for cosmic ray particles interacting resonantly
with Alfv\'en waves propagating parallel to the average magnetic field.
The transport equation can be derived from the Fokker-Planck equation by
a well-known approximate scheme (Jokipii 1966, Hasselmann \& Wibberenz 1968)
 which is commonly referred to as the diffusion-convection
equation for the pitch-angle averaged phase space density. Into the
equation the electromagnetic fields generated by MHD waves  enter through
the Lorentz force term. For fast ($v \gg V_{A}$) cosmic ray particles,
a vanishing cross helicity state of the Alfv\'en  waves and the power-law
turbulence spectrum
$$ g^{i}(k)=
(q-1)(\delta B_{i})^{2}k_{min}^{q-1}k^{-q}  \eqno(2.1)$$
the momentum diffusion coefficient

$$D_p={2\pi  (q-1)\over q(q+2)}
 \Bigl{(}{\delta B \over B } \Bigr{)} ^{2}
|\Omega| (k_{min})^{q-1}{V_{A}^{2} p^{2} \over v^{3-q}} \quad . \eqno(2.2)$$
Recently, Schlickeiser \& Miller (1997)
considered
cosmic ray particles interacting with oblique fast-mode waves propagating
in a low-$\beta$ plasm. In the cold
plasma limit the fast and slow magnetosonic waves merge to the fast-mode
waves with the same dispersion relation $\omega^{2}=V_{A}^{2}k^{2}$.
For such waves  they demonstrated that
the rate of adiabatic deceleration  vanishes, and  the
momentum diffusion coefficient  and spatial
diffusion coefficient $k_{\parallel}$ can be calculated as pitch angle
averages of the functions determined by the  non-vanishing Fokker-Planck
coefficient $D_{\mu \mu}$ and by $D_{pp}=\epsilon^{2}p^{2}D_{\mu \mu}(\mu)$
, where $\epsilon=V_{A}/v$.
Adopting isotropic fast-mode turbulence with a power-law turbulence
spectrum
they obtained
$$D_{\mu \mu} =
{\pi | \Omega | (q-1)(1-\mu^{2}) \over 4}\cdot $$
$$\cdot
\Bigl{(}{\delta B \over B } \Bigr{)} ^{2}
(k_{min}R_{L})^{q-1}[f_{T}(\mu)+f_{G}(\mu)] \quad ,
\eqno(2.3)$$
which is the sum of  transit-time damping
($f_{T}$) and gyroresonance interaction ($f_{G}$) contributions. The form of
$f_{T}$ admits the pitch angle scattering by transit-time damping of
super-Alfv\'enic
particles with pitch-angles contained in the interval  $\epsilon \leq \mid \mu
\mid \leq 1$. In the interval $\mid \mu \mid \le \epsilon$, where
no transit-time damping occurs, the gyroresonance ($n\ne 0$) interactions
provide a small
but finite contribution to the particle scattering rate. As a result
the momentum diffusion coefficient $D_p$ is predominantly determined by the
transit-time
damping contriburtion and the spatial diffusion coefficient is determined only by
the gyroresonance
interactions. Finally for the flat ($1 < q \leq 2$) and the steep ($q \ge 2$)
fast-mode turbulence  spectra (eq. 2.1) they obtained
$$D_{p}(1 <  q \leq 2) \simeq
{\pi (q-1)q \Gamma [q]\Gamma [2- (q/2)]
\over 2^{q+1}(4-q^2)\Gamma ^3[1+(q/2)]}\cdot$$
$$\cdot\Bigl{(}{\delta B \over B } \Bigr{)} ^{2} | \Omega | (R_{L}k_{min})^{q-1}
{V_{A}^{2}p^{2} \over v^{2} }\, ln {v \over V_{A}},  \eqno(2.4)$$
and
$$D_{p}(2 <  q \le 6) \approx
{\pi (q-1)(2q^{2}-3q+4) \over 16q(2q-3)}\cdot$$
$$\cdot\Bigl{(}{\delta B \over B}\Bigr{)} ^2 |\Omega| R_{L}k_{min}
{V_{A}^{2} p^{2}
\over v^{2}}\,
ln{v \over
 V_{A}} \qquad ,  \eqno(2.5)$$
respectively.

\section{Description of simulations}
The approach applied in the present paper is based on numerical Monte Carlo
particle simulations. The general procedure is simple:
test particles are injected at random positions into
a turbulent magnetized plasma and their trajectories are followed by
integrating the particle's equations of motion. Due to the presence
of waves, particles move diffusively in space and momentum.
By averaging over a large
number of trajectories one derives the diffusion coefficients for
turbulent wave fields. In the simulations we consider relativistic
particles with $v \gg V_{A}$ and use dimensionless units (cf Appendix A):
$\delta B \equiv \delta B / B_o$
for magnetic field perturbations, $1 / \Omega_o$ for time,
$k/k_{res}$ for wave vector and $p_o^2
\Omega_o$ for the momentum diffusion coefficient.

\subsection{The Wave Field Models}

In the modelling we consider a superposition of 384 MHD
waves propagating oblique to the
average magnetic field $\vec{B}_o \equiv B_o \hat{\bf e}_z$. The wave
propagation angle with respect to $\vec{B}_o$ is randomly chosen from a
uniform distribution within a cone (`wave-cone') along the mean field.
For a given simulation two symmetric cones are considered centered along
$\vec{B}_o$, with the opening angle $2\alpha$, directed parallel and
anti-parallel to the field direction. The same number of waves is
selected from each cone in order to model the symmetric wave field.
Related to the wave 'i' the magnetic field fluctuation vector $\delta
\vec{B}^{(i)}$ is given in the form:

$${\delta \vec{B}^{(i)} = \delta \vec{B}_o^{(i)} \sin (\vec{k}^{(i)}
\cdot \vec{r} - \omega^{(i)} t ) } \qquad . \eqno(3.1)$$
 
\noindent
The electric field fluctuation related to a particular wave is given
as ${\delta \vec{E}^{(i)}} = - {\vec{V}^{(i)}} \wedge {\delta \vec{B}^{(i)}}$ 
where $\vec{V}^{(i)}$ is a wave velocity.

For  Alfv\'en waves (A) we consider the respective dispersion relation

$$ \omega_A^2 = k_\|^2 V_A^2 \qquad , \eqno(3.2)$$

\noindent
where $V_A = B_o / \sqrt{4\pi \rho}$ is the Alfv\'en velocity in the
field $B_o$. The wave magnetic field polarization is defined by the
formula

$$ \delta \vec{B}_A = \delta B_A (\vec{k}, \omega_A) \, (\vec{k} \times
\hat{e} _z) \, k_\perp ^{-1} \qquad .  \eqno(3.3)$$

\noindent
In low-$\beta$ plasma  the fast-mode magnetosonic waves (M)
propagate with the Alfv\'en velocity and the respective relations are:

$$ \omega_M^2 = k^2 V_A^2 \eqno(3.4)$$

$$\delta \vec{B}_M = \delta B_M ( {\bf k} ,\omega_M) \, (\vec{k} \times
(\vec{k} \times \hat{e}_z)) \, k^{-1} \, k_{\perp}^{-1} \qquad .
\eqno(3.5)$$
\noindent
For all our simulations we adopt $V_{A}=10^{-3}c$.
One should be aware of the fact that the considered turbulence model is
unrealistic at large $\delta B$ and the present results can not be
considered as the exact ones. In particular, in the presence of a finite
amplitude turbulence the magnetic field pressure is larger than the mean
field pressure and the wave phase velocities can be greater than the
assumed here $V_A(B_o)$.

\subsection{Spectrum of the turbulence}
In our simulations we consider a power turbulence spectrum where
the irregular magnetic field,  obtained from the energy density which is
defined in equation (2.1), in the wave
vector range ($k_{min},k_{max}$) can be written

$$ \delta B(k)=\delta B(k_{min}) \Big( {k \over k_{min}}
\Big)^{- q /2}     \qquad \eqno(3.6)$$
where $k_{min}=0.08 (k_{max}=8.0)$ corresponds to the considered
longest (shortest) wavelength and $q$ is the wave spectral index.
In the present simulations we consider the flat spectrum with $q=1$ and
the Kolmogorov spectrum with $q=5/3$. We included the flat spectrum because
of
 our earlier simulations (Michalek \& Ostrowski 1996)
when we considered the  Alfv'en waves with $q=1$.
On the other hand  such kind of turbulence spectrum is very convenient for
numerical simulations due to presence of a big number of short waves.
For the flat spectrum wave vectors are drawn in a random way from the
respective ranges: $2.0 \le k \le 8.0$ for 'short' waves, $0.4 \le k \le 2.0$
 for 'medium' waves and $0.08 \le k \le 0.4$ for 'long' waves. The respective
 wave amplitude are drawn in a random manner so as to keep constant

$$ \big[ \sum_{i=1}^{384}(\delta B_o^{(i)})^2 \big]^{1/2} \equiv \delta
B , \qquad \eqno(3.7)$$
where $\delta B$ is a model parameter, and, separately in all mentioned
wave-vector ranges
$$ \big[ \sum_{i=1}^{128}(\delta B_o^{(i)})^2 \big]^{1/2} \equiv
 \big[ \sum_{i=129}^{256}(\delta B_o^{(i)})^2 \big]^{1/2} \equiv$$
 $$\equiv  \big[ \sum_{i=257}^{384}(\delta B_o^{(i)})^2 \big]^{1/2} \equiv
 {\delta B \over \sqrt{3}}.   \eqno(3.8)$$
Thus the wave energy carried  is uniformly divided into the all wave-vector
ranges.  As a second "realistic one" turbulence model we consider one with
 the Kolmogorov spectrum. The observed
spectra in the interplanetary space often take a such  form (Jokipii 1971).
In that case all wave vectors are drawn in a random manner from the whole
considered range  ($0.08 \le k \le 8.0$) but the amplitudes
$\delta B^{(i)}$
are fitted
according to Kolmogorov law (2.1) and to keep the formula (3.7).
In a such turbulence spectrum most of the energy is carried
by 'long' waves.
In the discussion below we will consider four different turbulence fields:
i. Alfv\'en waves with a flat spectrum - AF,         
ii. Alfv\'en waves with a Kolmogorov spectrum - AK,  
iii. Fast-mode waves with the flat spectrum - MF,    
iv.  Fast-mode waves with the Kolmogorov spectrum - MK.
\section{Results}
The results of the simulations are presented at the figures 1 and 2.
In Fig.~1 on the successive panels, the simulated momentum diffusion
coefficients $D_p$ for the  Alfv\'en and the magnetosonic turbulence
 are  presented.
 The derived
diffusion coefficients $D_p$ are given for different wave-cone opening
angles and for different turbulence amplitudes.
In  all cases a systematic increase
of $D_p$ with the amplitude
occurs, but the rate of this increase diminishes at larger non-linear
$\delta B$. We suspect the considered effect reflects partial particle
trapping in high-amplitude turbulence, decreasing randomness of the
particle momentum variation. There are no significant differences between
results obtained for both types turbulence spectrum, $q=1$ or $5/3$.
\begin{figure*}
\vspace{18cm}
\includegraphics{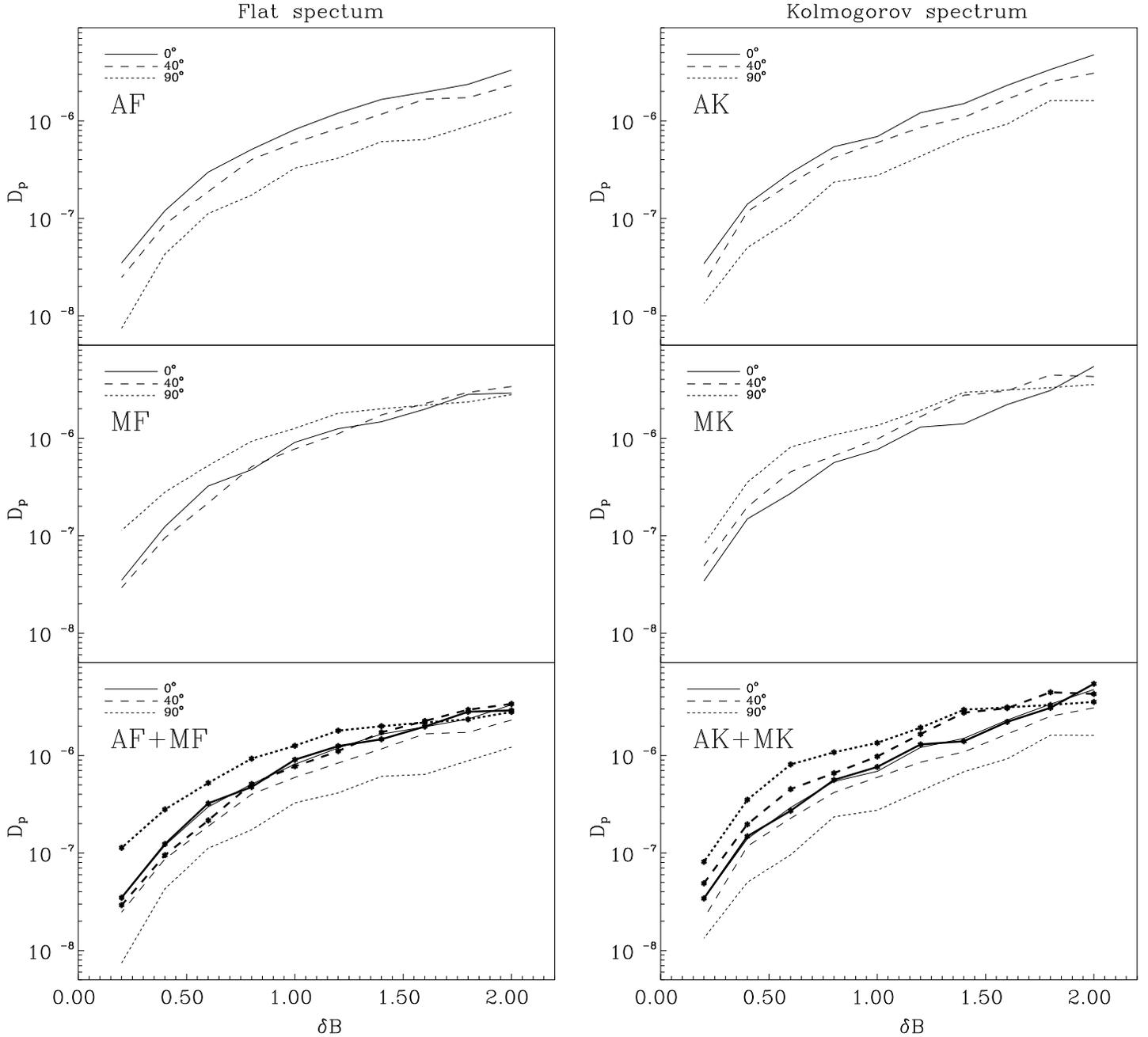}
\caption[]{Variation of the momentum diffusion coefficient $D_p$
versus the perturbation amplitude $\delta B$ and the wave propagation
anisotropy for the  flat spectrum and the Kolmogorov  spectrum. For
comparison results
for both the Alv\'en turbulence (thin lines) and the fast-mode turbulence
(thick lines with indicated simulation points) are superimposed  at the
lowest panels.}
\end{figure*}
\noindent
For the Alfv\'enic turbulence an increase of
the wave-cone opening angle $\alpha$ provides waves with smaller phase
velocities and leads to decreasing values of $D_p$.
Additionally, the values of the wave vector amplitude 
$k_{\parallel}$  decrease
leading to less efficient coupling mediated by  the cyclotron resonance.
 The trend is independent of the considered
 turbulence amplitude. In the case of fast-mode waves a more
complicated relation is observed. For the flat turbulence spectrum
 a small increase of the angle $\alpha$ leads to a slight decrease of $D_p$
 due to mentioned weakening of cyclotron resonance. However,
  the appearance of more
waves propagating nearly perpendicular to the ordered magnetic field
inverts this trend by allowing the transit time-damping resonance to become more
effective. For the Kolmogorov
turbulence an increase in opening angles is always followed by an increase
of $D_p$, but,   again reaches the maximum in the  presence of perpendicular
waves. The described
non-monotonic variations are clearly seen at low wave amplitudes in
Fig.~1, while for large amplitudes the phenomenon is less
pronounced. Generally, for large amplitudes of the magnetosonic waves the
value of $D_p$ does not strongly depend on the opening angle of the waves.
The simulation errors can be evaluated from comparison of the
results for $\alpha = 0^\circ$ of Alfven and fast mode waves, which should coincide. In Fig. 1 only the
clarity the results for $\alpha = 0^\circ$, $40^\circ$ and $90^\circ$
are presented, but the ones calculated by us for $\alpha = 30^\circ$ and
$60^\circ$ are consistent with the above description.

In order to explain this non-monotonic behavior one can relate to the
quasi-linear derivations of Schlickeiser \& Miller (1997).
For $n = 0$ the resonance condition for the transit-time damping may be
written as $v_\| = \omega / k_\|$ and, with the dispersion relation (4),
$V_A / v_\| = k_\| / k$. It is clear that for $V_A \ll v$ particles can
effectively interact with waves at wide range of $v_\|$ only if the
waves with $ k_{\perp} \gg k_\|$ are present. This fact explains
the effective transit-time damping interactions for magnetosonic waves
with isotropic space distribution of waves. Then,  the $D_p$ increase
results mainly from the presence of waves propagating
quasi-perpendicular to the mean magnetic field. In
our simulations we adopted
highly relativistic particles with $v=0.99c \gg V_A$, so that
this effect is the most pronounced.

In order to verify this
behaviour in more detail  we considered the interaction of relativistic
particles with waves propagating in narrow cones with
respect to
the ordered magnetic field. We performed simulations for cones $0^{o}\pm 5^{o},
45^{o}\pm5^{o}$,  $85^{o}\pm 5^{o}$ and results are shown at Fig.~2.
Only the presence of waves propagating nearly perpendicular
to the mean
magnetic field leads only to significant increase of $D_p$. For these
waves the transit-time damping resonance is responsible for the wave particle
coupling, while the waves propagating
at small angles or parallel to the average magnetic field only contribute
to the gyroresonance interaction.

\begin{figure}
\vspace{11cm}
\includegraphics{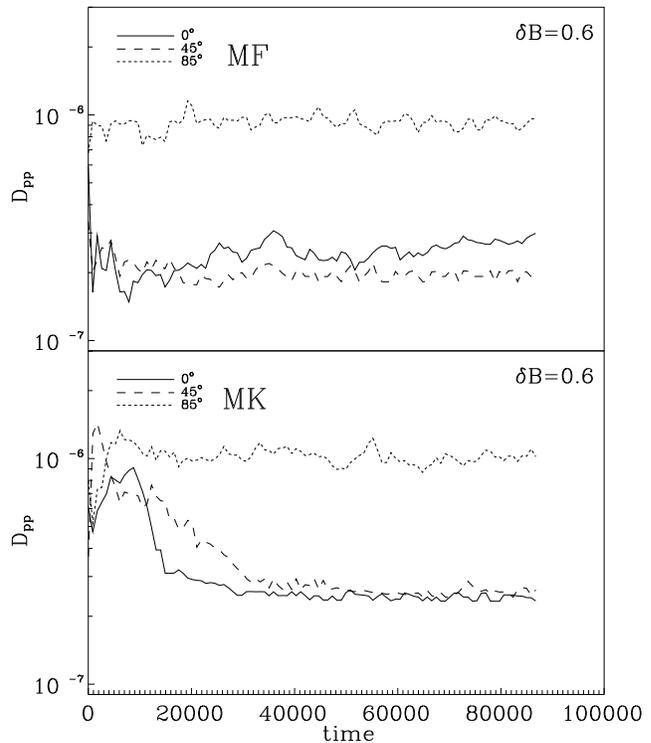}
\caption[]{Examples of the simulated $D_p$ for anisotropic
waves propagating close to the chosen pitch angle ($0^{o},45^{o},85^{o}$)
for both types of turbulence spectrum and $\delta B=0.6$.}
\end{figure}
\noindent

\section{Discussion and summary}
We considered momentum diffusion coefficient $D_{p}$ in the presence of
oblique Alfv\'enic and magnetosonic turbulence with amplitudes ranging
from small ones up to highly non-linear ones. The influence of the
degree of
anisotropy and of the waves spectrum slope on $D_{p}$ was also studied.
Generally, in all cases a systematic increase of $D_{p}$ with the wave
amplitude is observed. $D_{p}$ does not strongly depend on the type of
turbulence spectrum.
We confirm a substantial increase of $D_{p}$ for the fast-mode waves
in comparison to the Alv\'en waves of the same amplitude if the nearly
perpendicularly propagating  waves are included. That increase is caused by the
transit-time damping which occurs in the presence magnetosonic of waves
containing compressive component of $\delta \vec{B}$. In the presence
of a small amplitude magnetic turbulence a significant increase of $D_{p}$,
on a factor of 10, is achieved. This agrees well with the estimate of
Schlickeiser and Miller (1997) that the coefficient
  $D_p$
for the isotropic Kolmogorov-type spectrum in the presence of
magnetosonic waves is
about a factor $ln{V \over V_A}=$o$(10)$ larger than the one for the
 parallel-propagating Alfv\'en waves.
 In the case of  our simulations  for the Kolmogorov spectra this value
 is nearly constant for a wide range of waves' amplitudes and 
 for 
  isotropic wave  spectrum  our results
are consistent with the quasilinear estimates. The approximation
worsens for the largest non-linear amplitudes ($\delta B/B_0)>1.5$)
when the considered ratio diminishes on about $50\%$.

We also found that for large
amplitudes of magnetosonic waves the difference between $D_p$ does not
strongly depend on the opening angle of the waves. When the strength of the
compression is high enough  most particles can be reflected
before complete penetrating the region of the compression they  could
only be accelerated due to a simple second order Fermi mechanism. As the
amplitude decreases, the number of particles that are reflected decrease.
For very small amplitudes particles can be reflect at the wave when their
parallel velocities in the wave frame is about zero. That occurs when
$v_{\parallel}=\omega/k_{\parallel}$ what is condition for the transit-time
damping. Hence, that process could be called small-amplitude Fermi acceleration
(Achterberg 1981).  We should remember that
 the wave phase velocity are expected
to be larger in realistic turbulence than the assumed here $V_{A}(B_{o})$.
Therefore our results for large $\delta B$ should be consider rather as
the lower limits.

Considerations of the MHD waves
propagation oblique to the mean
magnetic field (e.g. Tademaru 1969, Lee and Volk 1975) show
that such waves are subject to effective processes
dissipating their energy. Because of that the considered here effects of
the momentum diffusion enhancement due to fast-mode waves can occur only
in a volume with acting the turbulence generation force. For example, in
vicinity of the strong shock or in a region of the magnetic field
reconnection the required fast-mode perpendicular waves are expected to be
effectively created.

\begin{acknowledgements}
MO thanks to Prof. Richard Wielebinski for a kind invitation to
Max-Planck-Institut f\"ur Radioastronomie in Bonn and for hospitality.
A part of the present work was done during this visit.
He is also grateful to Hui Li for a useful discussion.
GM \& MO acknowledge the Komitet Bada\'n Naukowych support through the grant PB 179/P03/96/11.
\end{acknowledgements}

\section{References}

Achterberg A.,  1979, A{\&}A,  76, 276 

Achterberg A.,  1981, A{\&}A, 97,259

Blandford R.D., Rees M.I., 1978, Physic Scripta, 17, 3543

Burn B.I., 1975, A{\&}A, 45,435.

De Young D.S., 1976, Ann. Rev. Astron. Astrophys., 14,447

Eilek J.A., 1979, Astrophys. J., 230, 373

Hall D.E., Sturrock P.A., 1967, Phys. Fluid, 10,2620

Hasselmann K., Wibberenz G., 1967, Z. Geophys., 34,353

Jokipii J.R., 1966, Astrophys. J., 146,480

Jokipii J.R., 1971, Reviews of Geophysics and Space  Physics,9, 27

Kulsrud R.M. Ferrari  A., 1971, Astrophys. Space. Sc. 12,302

Lee M.A., Volk H.J., 1975, Astrophys. J., 198, 485

Micha\l ek G., Ostrowski M.,  1996, Nonlinear Processes in Geophysics 3, 66 

Micha{\l}ek G., Ostrowski M., 1997, A{\&}A (submitted) 

Miller J.A., LaRosa T.N., Moore R.L., Astrophys. J., 1996, 461, 445

Schlickeiser R., 1989, Astrophys. J., 336, 243

Schlickeiser R., Miller J.A., 1997, submitted. 

Stix T.H., 1962, The Theory of Plasma Waves McGraw-Hill, New York

Tademaru E., 1969, Astrophys. J., 158,.958

\appendix
\section{Summary of notation}

$a_{2} \equiv D_{p}$-- momentum diffusion coefficients

${\bf B} = {\bf B}_0 + \delta {\bf B}$  -- magnetic induction vector 

${\bf B}_0$ -- regular component of the background magnetic field

($B_0 = 1$ in the simulations) 

$\delta {\bf B}$ -- turbulent component of the magnetic field 

$c$ -- light velocity ($c = 1$ in simulations)  

$D_{\mu \mu}$-- Fokker-Planck coefficient

${\bf E}$ -- electric field vector  

$g^{j}(k)$ -- magnetic energy density for given wave

$e$ -- particle charge 

$k_{res}=2 \pi/r_g$

$k_{\parallel}$ -- wave vector along mean magnetic field

${\bf p}$ -- particle momentum vector  

$q$ -- spectral index of waves

$r_g$ -- particle gyro-radius  

$R_{L}={v \over |\Omega|}$ -- particle Larmour radius

${\bf v} \equiv c^2 {\bf p} / \varepsilon $ -- particle velocity vector

$ V_A $ -- the Alfv\'en velocity in the field $B_0 $  

$v_{\parallel}$ -- velocity along mean magnetic field

$\alpha$--openin angle of MHD waves

$\gamma \equiv (1-v^2/c^2)^{-1/2} $ -- the Lorentz factor  

$i$ -- number of give waves

$k$ -- wave-vector 

$\kappa_ \perp $ -- transverse (cross-field) diffusion coefficient  

$\kappa_\|$ -- parallel diffusion coefficient  

$m$ -- particle mass ($m = 1$ in simulations)  

$\omega$ -- wave frequency   

$\Omega \equiv e B / \gamma m c$  -- particle angular velocity 

$\Omega^{(o)} \equiv e B_o / \gamma m c$ 

$\Theta$  -- the momentum pitch-angle with respect to ${\bf B}_0$ 

($\mu \equiv \cos \Theta$)                                     

$\varepsilon$ -- particle energy  

$\epsilon={V_{A}\over v}$

\end{document}